\documentclass[twoside]{article}[15pt]

\usepackage[dvips]{epsfig}
\usepackage{rotating}

\begin{document}

\large
\markboth{ Li et al.}{isochores merit the prefix 'iso'}

\begin{center}

{\huge
Isochores Merit the Prefix `Iso'
}

\vspace{0.4in}
{\Large
Wentian Li$^{1,*}$, 
Pedro Bernaola-Galv\'{a}n$^2$,
Pedro Carpena$^2$,
Jose L. Oliver$^3$ \\ 
}
\vspace{0.3in}

{\sl 1. Center for Genomics and Human Genetics, North Shore - LIJ Research Institute}\\
{\sl 350 Community Drive, Manhasset, NY 11030, USA}\\
{\sl 2. Departmenta de F{\'\i}sica Aplicada II, Universidad de M\'{a}laga,
E-39071, M\'{a}laga, Spain}\\
{\sl 3. Departmento de Gen\'{e}tica, Instituto de Biotechnolog{\'\i}a, 
Universidad de Granada, E-18071 Granada, Spain} 

\end{center}

\vspace{0.2in}

Email addresses: wli@nslij-genetics.org,
rick@uma.es, pcarpena@ctima.uma.es, oliver@ugr.es.

* The corresponding author.  

Running title: isochores merit iso

\vspace{1.4in}

{\bf Abbreviations and acronyms:}

ANOVA: analysis of variance; 
IHGSC: international human genome sequencing consortium;
MHC: major histocompatibility complex

\newpage

\begin{center}
{\bf Abstract}
\end{center}

The isochore concept in human genome sequence was challenged in
an analysis by the {\sl International Human Genome Sequencing Consortium} 
(IHGSC).  We argue here that a statement in IGHSC analysis concerning
the existence of isochore is incorrect, because it had applied 
an inappropriate statistical test. To test the existence of 
isochores should be equivalent to a test of homogeneity of 
windowed GC\%.  The statistical test applied in the IHGSC's 
analysis, the binomial test, is however a test of a sequence 
being random on the base level. For testing the existence of 
isochore, or homogeneity in GC\%, we propose to use 
another statistical test: the analysis of variance (ANOVA). 
It can be shown that DNA sequences that are rejected by binomial 
test may not be rejected by the ANOVA test.

%%%%%%%%%%%%%%%%%%%%%  section 1 %%%%%%%%%%%%%%%%%%%%%%%

\section*{Background}

\noindent

The degree of homogeneity in base composition in human genome 
is a fundamental property of the genome sequence. Not only 
does it characterize the organization and evolution of
the genome, but also it provides a context of many practical
sequence analysis. Statistical quantities such as GC\%, used 
for sequence analyses such as computational gene recognition, 
should be sampled from a homogeneous region of the sequence. 
If these quantities are sampled from an inhomogeneous region, 
error is introduced and the quality of a sequence analysis such 
as the performance of gene prediction, could be affected.

It has been known for a long time from the work of Bernardi's 
group that there are compositional homogeneous regions in human 
genome with sizes of at least 200-300 kb \cite{bernardi93, bernardi95}. 
These homogeneous regions are  called ``isochores" \cite{cuny}, 
and the whole genome is a mosaic of isochores. Recently, however,
this view of human genome is questioned in an initial
analysis of human genome draft sequence \cite{lander}. 
The analysis presumably shows that no sequence of 300-kb length examined 
could be claimed to be homogeneous (``... the hypothesis of 
homogeneity could be rejected for each 300-kb window in the 
draft genome sequence", page 877 of \cite{lander}, and 
a stunning statement was made that, essentially, isochore concept does 
not hold (``... isochores do not appear to merit the prefix `iso'", 
page 877 of \cite{lander}).

The purpose of this Letter is to show that an incorrect
statistical distribution for windowed GC\% is assumed in 
\cite{lander}, based on an unrealistic condition
for DNA sequences. As a result, the statistical test used in 
\cite{lander} is invalid. We will present a 
correct statistical test, assuming a more reasonable
statistical distribution of windowed GC\%. Under the new
test, the conclusion concerning the existence of isochore
is drastically altered. Although our testing result may still
depend on the window size at which GC\% is sampled, and
may possibly depend on the choice of GC\% groups, 
it is clear that the test in \cite{lander} is too biased 
towards rejecting the homogeneity null hypothesis, 
and sequences that fail the test in \cite{lander}
usually do not fail our new test.

\section*{Results}

\noindent

For a sequence to be homogeneous in GC\%, the mean/average 
of windowed GC\% values sampled from one region of the sequence 
should be similar to that in another region, with a consideration
on the amount of allowed variance.  In other words, to 
claim that a sequence is homogeneous, not only do we need 
to calculate means of GC\% along the sequence, but also 
we need to know the variance. Generally speaking, the
mean and the variance are two independent parameters of a 
statistical distribution. However, for the homogeneity test in 
\cite{lander}, the variance is assumed to be a function of the 
mean, thus it is not independently estimated. 

In \cite{lander}, the windowed GC\% is assumed 
to follow a binomial distribution. For a binomial
distribution to be true, bases within the window should be 
uncorrelated, similar to tossing a coin many times.
Violating this assumption invalids the use of binomial
application. The more reasonable statistical distribution 
of GC\% should be the normal distribution which, 
unlike the binomial distribution, has two independent 
parameters (mean and variance). Mean value can be 
estimated from a window, whereas variance can
be estimated from a group of windows.

To illustrate our point, we analyze two well known isochore 
sequences, the Major Histocompatibility Complex (MHC) 
class III and class II sequences on human chromosome 6 
\cite{fukagawa95,fukagawa96,stephens,beck}), 
with lengths 642.1 kb and 900.9 kb, respectively. The
exact borders of the two isochore sequences are determined
by a segmentation procedure \cite{oliver01,li01}
and an online resource on isochore mapping \cite{web_ugr}). 
We first repeat the test in \cite{lander} that these two 
sequences, when viewed as a collection of many 20 kb windows, are sampled 
from a binomial distribution. According to  \cite{lander},
a rejection of this test is considered  to be an evidence for
heterogeneity.  The test results are included in Table 1, which 
clearly shows that the variances of GC\% values sampled from 20-kb 
windows are much larger than expected from a binomial distribution,
with $p$-value close to be 0 ($< 10^{-50}$).

This result, that the variance of GC\% sampled from windows
is much larger than expected by binomial distribution, has
been known for a long time 
\cite{sueoka59,sueoka62,cuny,li98}, \cite{clay} (and the 
references therein).  It is not surprising that the 
binomial distribution assumption is rejected even for 
isochore sequences as shown in Table 1.  Nevertheless, 
this rejection only shows that a 20-kb window is not 
a series of 20000 uncorrelated bases; it is not a rejection
of homogeneity of windowed GC\% along the sequence.

To reaffirm our belief that the binomial test used in
\cite{lander} is a test of randomness of the
sequence instead of homogeneity, one bacterial sequence 
({\sl Borrelia burgdorferi}, 910.7 kb) and two randomly generated 
sequences (with same length and base composition as the MHC 
class III and class II sequences) are used for test. Table 1 
shows that the null hypothesis cannot be rejected by the binomial 
test for the two random sequences, but it is rejected for 
the  {\sl Borrelia burgdorferi}, a particularly homogeneous 
genome, as shown in a recent survey of archaeal and 
bacterial genome heterogeneity \cite{bern02}.

We would like to suggest that the more reasonable statistical 
distribution of windowed GC\% is the normal/Gaussian 
distribution, and the more appropriate test of homogeneity 
of these GC\% values along a sequence is the analysis of 
variance (ANOVA). There are at least two reasons to believe 
that ANOVA is the more appropriate test. First, it is a 
test of equality between means, which is identical to 
the intuitive meaning of homogeneity, i.e., GC\% are the
same along the sequence. Second, ANOVA and normal distribution 
reflects the real situation of DNA sequences that these are
not random sequences, and windowed GC\%'s exhibit higher 
values of variances.  ANOVA allows the variance to be 
estimated from the data, rather than being fixed by 
the mean value as in binomial distribution.
ANOVA was previously applied to the study of inter-chromosomal
homogeneity of yeast genome 
\cite{li98,oliver98}.

To apply ANOVA to test homogeneity, we split a sequence into
several super-windows, and several windows per super-window. 
GC\% from each window is calculated. The null hypothesis is 
that the mean of windowed GC\%'s in each super-window is the 
same. The simplest selection of super-windows and windows is to 
assume all windows to have the same length. To match the 
discussions in \cite{lander}, we choose 20-kb windows 
and 300-kb super-windows. This corresponds to roughly 2 
super-windows, 16 windows per super-window for the MHC class 
III sequence, and 3 super-windows, 15 windows per super-window 
for the MHC class II sequence. ANOVA test results of these 
two isochores are listed in Table 2. The $p$-values are 
0.192 and 0.323, respectively, for MHC class III and class 
II sequence. The null hypothesis, that means of GC\% in 
different super-windows are the same, is not rejected.

When the ANOVA test is applied to the {\sl Borrelia burgdorferi}
genome sequence and two randomly generated sequences, null
hypothesis cannot be rejected, indicating that all three 
sequences are homogeneous at the respective window and
super-window sizes (20 kb and 300 kb). This is a more satisfactory
situation than the binomial test because now a homogeneous
bacterial sequence is indeed confirmed to be homogeneous
by the test.

\section*{Discussions}

\noindent

Due to the ``domains within domains" phenomenon in DNA sequences
\cite{li94,bern96,li97}, we should not assume automatically 
that a homogeneity test result obtained at 20-kb window 
and 300-kb super-window will hold true for other window 
and super-window sizes. To check this, we carry out ANOVA tests 
on the MHC class III and class II sequences at other window 
and super-window sizes.  Fig.1 shows the result for the ANOVA test 
result ($-\log_{10}(p-$value) ) for window sizes of around 20 kb, 
10 kb, 5 kb, and 2.5 kb, and the sequence is partitioned into 
2, 3, 5, 8 (2,3,5,9) super-windows for MHC class III (II) sequence.

Several observations could be made from Fig.1. First, when GC\%'s
are sampled from (e.g.) 20-kb windows, changing the number of super-windows
(i.e. number of partitions of the sequence) does not greatly
influence the ANOVA test result. This change corresponds to
a regrouping of windowed GC\%'s. Generally speaking, if the sequence 
is homogeneous with all GC\% values (taken from a fixed window 
size) having the similar value, regrouping these values does 
not make an insignificant result to be significant.

Second, the ANOVA test becomes more significant when the 
window size decreases. This observation is understandable 
because at smaller length scales, GC\% fluctuations are no
longer averaged out. These smaller-length-scale fluctuations 
could be due to repeats, insertions, foreign elements, etc. 
For MHC class II sequence, as the subwindow size is reduced 
to around 2.5 kb, the ANOVA test result is typically significant 
(Fig.1).  This is consistent to the definition of isochores as 
``fairly homogeneous" (as versus ``strictly homogeneous")
segments above a size of 3 kb \cite{bettecken,bernardi00}, 
and justifies the ``coarse graining" procedure 
to locate isochore boundaries in \cite{oliver01}.

Third, two isochore sequences may look similar at one length scale
(e.g. 20 kb), but quite different at another length scale. Fig.1
shows that MHC class II sequence is more heterogeneous than MHC 
class III sequence when viewed at the 2-10 kb length scales.
It is known that GC-poor sequences are generally considered 
to be more homogeneous than GC-rich sequences, or more accurately, 
a sequence with a GC\% closer to 50\% is more heterogeneous 
than a sequence whose GC\% is far away from 50\% 
\cite{sueoka62,cuny,clay}. Since the GC\% of MHC 
class III and II sequence is 51.9\% and 41.1\%, respectively,
we might expect MHC class II sequence to be more homogeneous
than class III sequence. Interestingly, Fig.1 shows the contrary.

To conclude, the binomial test used in \cite{lander} should 
not be a test of homogeneity if the expected variance
does not reflect the true variance in the sequence. The reason
that the expected variance in a binomial test (which is
derived from the mean GC\% instead of being an independent
parameter)  is unrealistic is because the underlying base 
sequence is not random/uncorrelated. We are naturally led to 
the ANOVA test if we actually estimate the variance from the 
data.  With ANOVA tests, it is clear that homogeneous regions 
of GC\% in human genome do exist; in other words, isochores exist.

\large

\section*{Methods}

\noindent

{\bf Binomial test:} Following \cite{web_nature}, 
a binomial test is applied to many GC\% values measured 
from a fixed-sized window (e.g. 20 kb). For example, 
if the sequence length is 900 kb, there are $n=$45 such 20-kb
windows and 45 GC\% values. The variance of these GC\%'s ($\sigma^2$) is calculated, 
and the variance as expected from a binomial distribution is 
$\sigma^2_0= m(1-m)/20000$, where $m$ is probability of G or C. 
The value of $m$ can be estimated  by the actual GC\% of the 
sequence. The test statistic is $c^2 = (n-1) \sigma^2/\sigma^2_0$. 
For null hypothesis (that windowed GC\% measurements do follow 
binomial distribution, which is true when the underlying 
base sequence is random/uncorrelated within the window), 
$c^2$ follows the $\chi^2_{df=n-1}$ distribution 
(e.g. $\chi^2_{df=44}$ in our example). For any given $c^2$
value, the $p$-value can be determined by the corresponding
$\chi^2$ distribution.

{\bf ANOVA test:} ANOVA test (analysis of variance) is applied
to several groups of GC\%'s (as a comparison, binomial test is
only applied to one group of GC\%'s). The concept of ``group" and
``member" in ANOVA now becomes ``super-window" and ``window" here.
The number of super-windows partitioned in a sequence is $a$,
and the number of windows in the super-window $i$ is $n_i$. 
The two ``sum of squares" (SS) are defined:
SS$_w= \sum_{i=1}^a \sum_{j=1}^{n_i} (GC\%_{ij} -\overline{GC\%_i})^2 $ 
(within a group), and  
SS$_a = \sum_{i=1}^a n_i (\overline{GC\%}_i - \overline{\overline{GC\%}})^2$ 
(among groups). The test statistic is $F=SS_a/SS_w \times \sum_{i=1}^a (n_i-1)/(a-1)$.
The distribution of $F$ under null (i.e., GC\%$_1$=GC\%$_2$= 
$\cdots$ GC\%$_a$) is known, and this distribution can be used 
to determined the $p$-value.

\section*{Acknowledgments}

\noindent

We would like to acknowledge the financial support from the 
5th Anton Dohrn Workshop at Ischia (2001) where some of the 
ideas presented here were discussed.  W.L. acknowledges 
partial support from NIH contract N01-AR12256, P.B.G., 
P.C. and J.L.O. acknowledge the grant support BIO99-0651-CO2-01 
from the Spanish Government.

\newpage

\begin{figure}[t]
\begin{center}
  \begin{turn}{-90}
  \epsfig{file=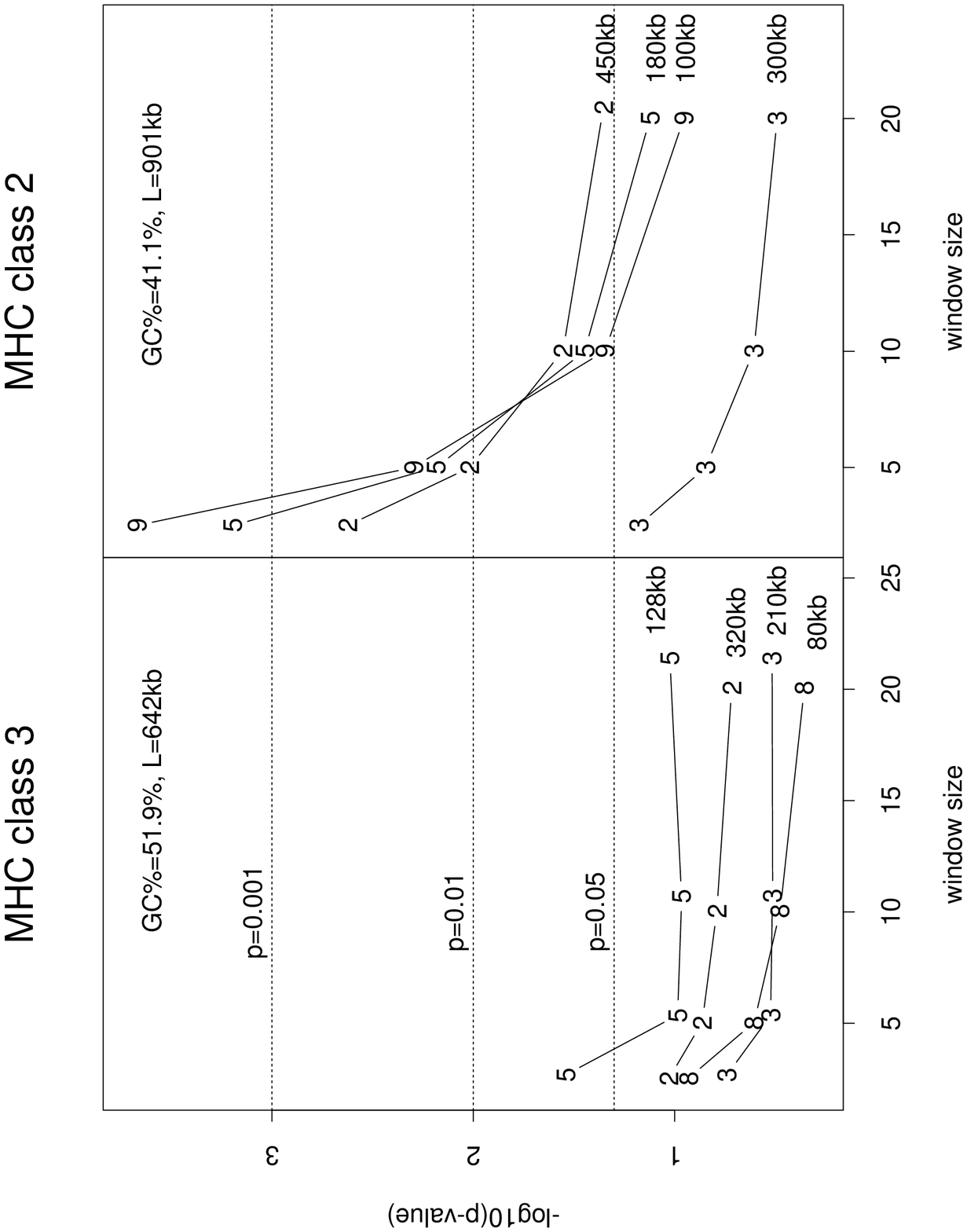, height=16cm, width=9cm}
  \end{turn}
\end{center}
\caption{
The $-\log_{10}(p-$value) of ANOVA tests as a function of the window 
sizes, for MHC class III (left) and MHC class II (right) 
sequences. These tests with the same number of super-windows 
are connected in a line. The size of the super-window 
and the number of super-windows in the sequence is 
indicated for each line.
}
\label{fig1}
\end{figure}

\begin{table}
\begin{center}
\begin{tabular}[t]{cccccccc}
\hline
seq & \# win ($n$) & mean & var $\sigma^2$ & binomial var $\sigma^2_0$
 & $\sigma^2/\sigma^2_0$ &  $c^2= (n-1) \sigma^2/\sigma^2_0$ & $p$-value \\
\hline
MHC class III & 32&  0.5188 & 0.0005345 & 0.00001248 & 42.8215 & 1327.47 & {\bf 0} \\
MHC class II & 45 & 0.4105 & 0.0007268 & 0.00001210 &  60.0709 & 2703.19 & {\bf 0} \\
\hline
random (class III) & 32 & 0.5185 & 0.00001137 & 0.00001248 & 0.9110 & 28.2402
 & {\sl 0.609} \\
random (class II) & 45 & 0.4106 & 0.00001255 &0.00001210 & 1.0369
 & 45.6244 & {\sl 0.404} \\
\hline
B. burgdorferi & 45 & 0.2859 & 0.0001515 & 0.00001021 & 14.8432
 & 653.099 & {\bf 0} \\
\hline
\end{tabular}
\end{center}
\caption{
{\bf Testing the hypothesis that GC\% values sampled from 20-kb windows 
follow a binomial distribution}. Five sequences are tested: MHC class III
and MHC class II isochore sequences, two random sequences similar these
two MHC sequences (same length and same base composition), and bacterium
{\sl Borrelia burgdorferi} genome sequence. Detailed explanation
of column headers: 
1. Sequence name.
2. Total number of windows in the sequence ($n$), with each contributing a GC\% value.
3. Mean of the GC\% ($m$).
4. Variance of the GC\% ($\sigma^2$).
5. Variance of GC\% expected from a binomial distribution
($\sigma_0^2 = m(1-m)/20000$).
6. Ratio of the two variances $\sigma^2/\sigma^2_0$.
7. test statistic $c^2= (n-1) \sigma^2/\sigma^2_0$.
8. $p$-value from the binomial distribution test.
}
\end{table}

\begin{table}
\begin{center}
\begin{tabular}[t]{cccccc}
\hline
 & df & SS & MS & F-value & $p$-value \\
\hline
\multicolumn{6}{l}{MHC class III (sw=2, w=16)}\\
\hline
between windows & 1 & 0.0009159 & 0.0009159 & 1.781 & {\sl 0.192}  \\
within windows & 30 & 0.01543 & 0.0005143 & &  \\
\hline
\multicolumn{6}{l}{MHC class II (sw=3, w=15)}\\
\hline
between windows & 2 & 0.001658  & 0.0008288 & 1.162 & {\sl 0.323} \\
within windows & 42 & 0.02997 & 0.0007137 & &  \\
\hline
\multicolumn{6}{l}{random seq similar to class III (sw=2, w=16)}\\
\hline
between windows & 1 & 0.00000288 & 0.00000288 & 0.247 & {\sl 0.623 }  \\
within windows & 30 & 0.0003496 & 0.00001165 & &  \\
\hline
\multicolumn{6}{l}{random seq similar to class II (sw=3, w=15)}\\
\hline
between windows & 2 & 0.00004546 & 0.00002273 & 1.884 & {\sl 0.165 } \\
within windows & 42 & 0.0005066 & 0.00001206 & &  \\
\hline
\multicolumn{6}{l}{B. burgdorferi (sw=3, w=15)}\\
\hline
between windows & 2 & 0.0002064& 0.0001032 & 0.671 & {\sl 0.517 } \\
within windows & 42 & 0.006461& 0.0001538& &  \\
\hline
\end{tabular}
\end{center}
\caption{
{\bf ANOVA test results of the five sequences}
(two MHC isochore sequences and their randomized sequences, 
and bacterium {\sl Borrelia burgdorferi} sequence). 
$df$: degrees of freedom; $SS$: sum of squares. 
$MS$: mean squares. $F$-value: test statistic value; 
$p$-value: $p$-value from the ANOVA test. $sw$ and $w$ are the
number of super-windows and windows.
}
\end{table}

\end{document}